\def\hmpc{\,{h^{-1} {\rm Mpc}}}
\documentclass[usenatbib]{mn2e}
\usepackage{epsfig}
\usepackage{amssymb}
\usepackage{amsmath}
\usepackage{textcomp}
\usepackage{longtable}

\begin{document}

\title[Cosmic velocity field]{%
Velocities hasten to tell us about the Universe}

\author[Ma \& Scott]{Yin-Zhe Ma$^{1,2,\dagger}$ \& Douglas Scott$^{1,\star}$\\
$^1$Department of Physics and Astronomy, University of British Columbia,
 Vancouver, V6T 1Z1, BC Canada.\\
$^2$Canadian Institute for Theoretical Astrophysics, Toronto, M5S 3H8,
 Ontario, Canada.\\
emails: $^{\dagger}$mayinzhe@phas.ubc.ca;\,
$^{\star}$dscott@phas.ubc.ca}

\maketitle

\begin{abstract}
The peculiar velocities of galaxies are driven by gravity, and hence
hold the promise of probing details of how gravity forms structures.   In
particular it is possible to constrain
cosmological parameters and to test extensions to the standard model,
such as modifications to the theory of gravity or the existence of
primordial density perturbations which are non-Gaussian.  This constraining
power has been frustrated by systematic effects, but we appear to be
entering an era when velocity measurements may finally be living up to
their promise.
\vskip 0.3 truein
\end{abstract}

\section{Introduction}
\label{sec:intro}

Gravity grows galaxies from low contrast seed perturbations, and hence
the galaxies will have motions which depend on the
neighbouring density field.  It has been understood since at least the
1960s (e.g.\ in the work of van Albada) that so-called ``peculiar
velocities'' could help us to understand structure formation.  In the
1970s the theoretical framework for such calculations was worked out
(by Peebles and others) and by the 1980s there was a vigorous programme
of estimating velocities and using them to constrain cosmologies.

The cosmic velocity field appeared to yield a high average
density for the Universe and hence provided support for the ``Standard
Cold Dark Matter'' paradigm \citep[e.g.,][]{Dekel93}.
However, it became clear that the determination
of redshift-independent distances (necessary to estimate the velocities)
was fraught with problems, leading to correlated errors that were
fiendishly difficult to get rid of.  Hence, by the 1990s, the use of
velocities was largely abandoned in favour of other cosmological probes.

However, the situation now seems to be changing for the better, with new
data sets and techniques enabling the cosmic flow field to be mined for
cosmological information.  This is a gratifying development, since velocities
offer some distinct advantages as tests of the cosmological paradigm in which
gravity plays a dominant role on large scales.

\section{Testing the standard model}
\label{sec:testing}

In the Universe, galaxies are not only moving outwards with
cosmic expansion (known as the Hubble flow), but also have their
own peculiar motions, coming from initial perturbations enhanced by gravity.
The observed galaxy redshift is the combined effect of cosmic expansion
and the radial component of these peculiar velocities \citep{Peebles93}:
\begin{equation}
cz=H_{0}r+ \hat{r}\cdot \vec{v}(\vec{r}) , \label{eq:cz}
\end{equation}
where $z$ and $r$ are the redshift and true distance of the
object, $H_{0}$ is the Hubble constant
($=100h\,{\rm km}\,{\rm s}^{-1}\,{\rm Mpc}^{-1}$)
and $\vec{v}(\vec{r})$ is the
proper motion of the galaxy (at position $\vec{r}$) with respect
to a comoving frame (such as the cosmic microwave
background, or CMB, rest frame).  This equation shows that only the
radial component of a galaxy is directly measurable.

Understanding the motion of galaxies and galaxy clusters is an
important way of exploring the large-scale nature of the
Universe.  Any viable model should not only be
able to predict the evolution of the density field, but also
the velocity field and correlations between the two.
Therefore, one can use maps of the peculiar
velocity field to determine the validity of underlying cosmology models,
making it a powerful way of testing structure formation theories.
Peculiar velocity studies can be used to test basic assumptions such as
the ``Copernican'' principle of cosmology, i.e.\ whether we are in a
special position in the Universe, and how far we have to look to see
beyond our inhomogeneous patch.  We can also investigate fundamental
issues, such as testing modified theories of gravity
\citep[see e.g.][]{Hellwing} or constraining the non-Gaussianity of
primordial density perturbations \citep[see e.g.][]{MaTS}.

\section{Velocity field versus density field}
\label{sec:velocity-density}

\begin{figure}
\centerline{
\includegraphics[bb=0 0 587 297, width=3.2in]{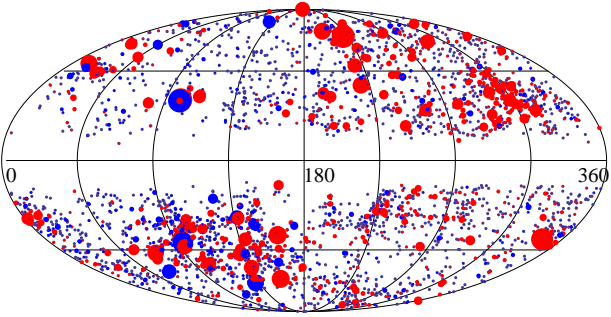}}
\caption{Peculiar velocity catalogue of 3456 spiral field galaxies, selected in
the \textit{I}-band and plotted in Galactic coordinates (taken from
\citealt{Ma12a}).  The red points are moving away from us and the blue
points are moving towards us, with the
size of the points proportional to the magnitude of the
line-of-sight peculiar velocity.}
\label{fig:sfifig}
\end{figure}

\begin{figure*}
\centerline{
\includegraphics[bb=0 0 626 416, width=3.2in]{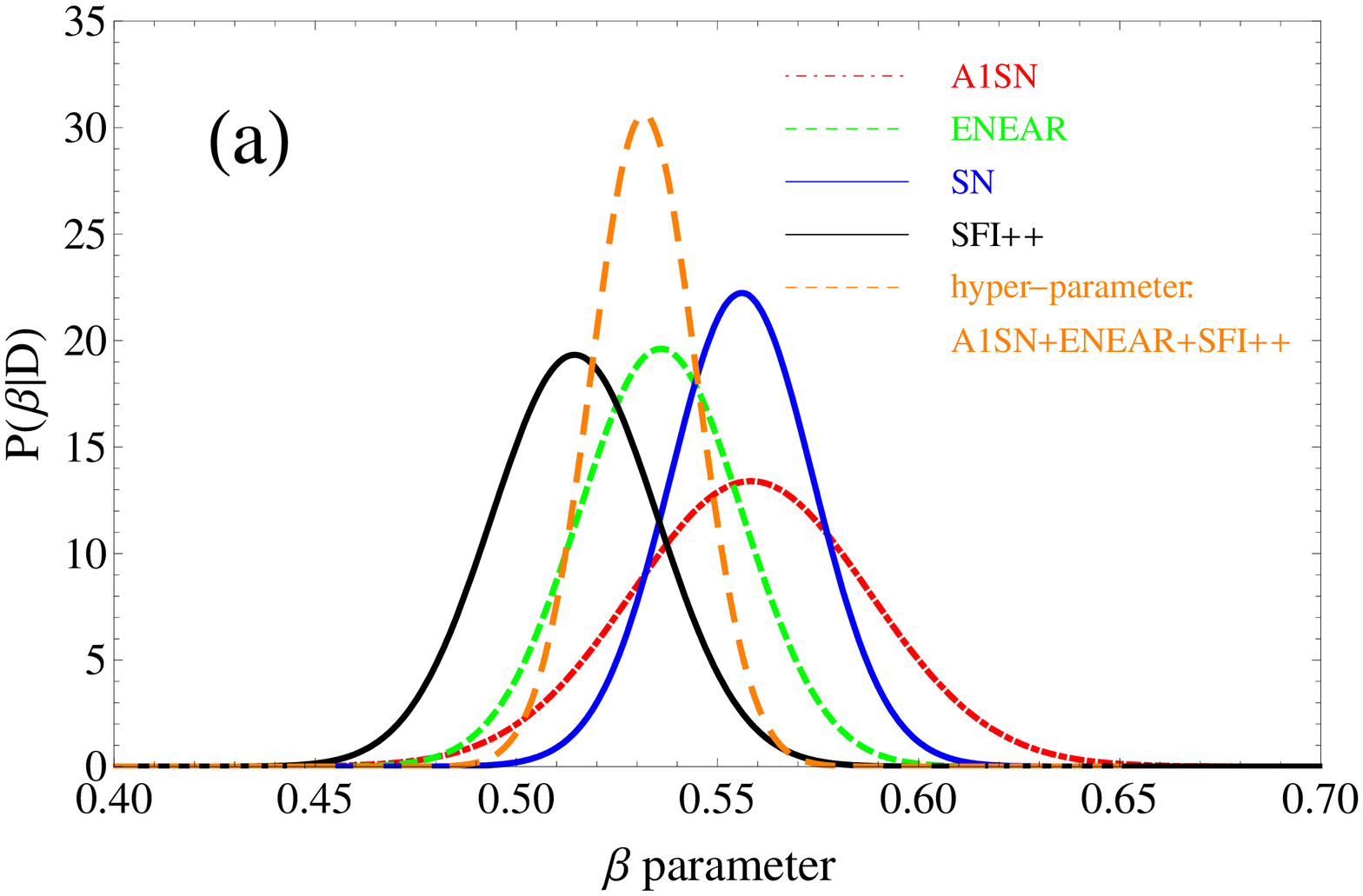}
\includegraphics[bb=0 0 671 435, width=3.2in]{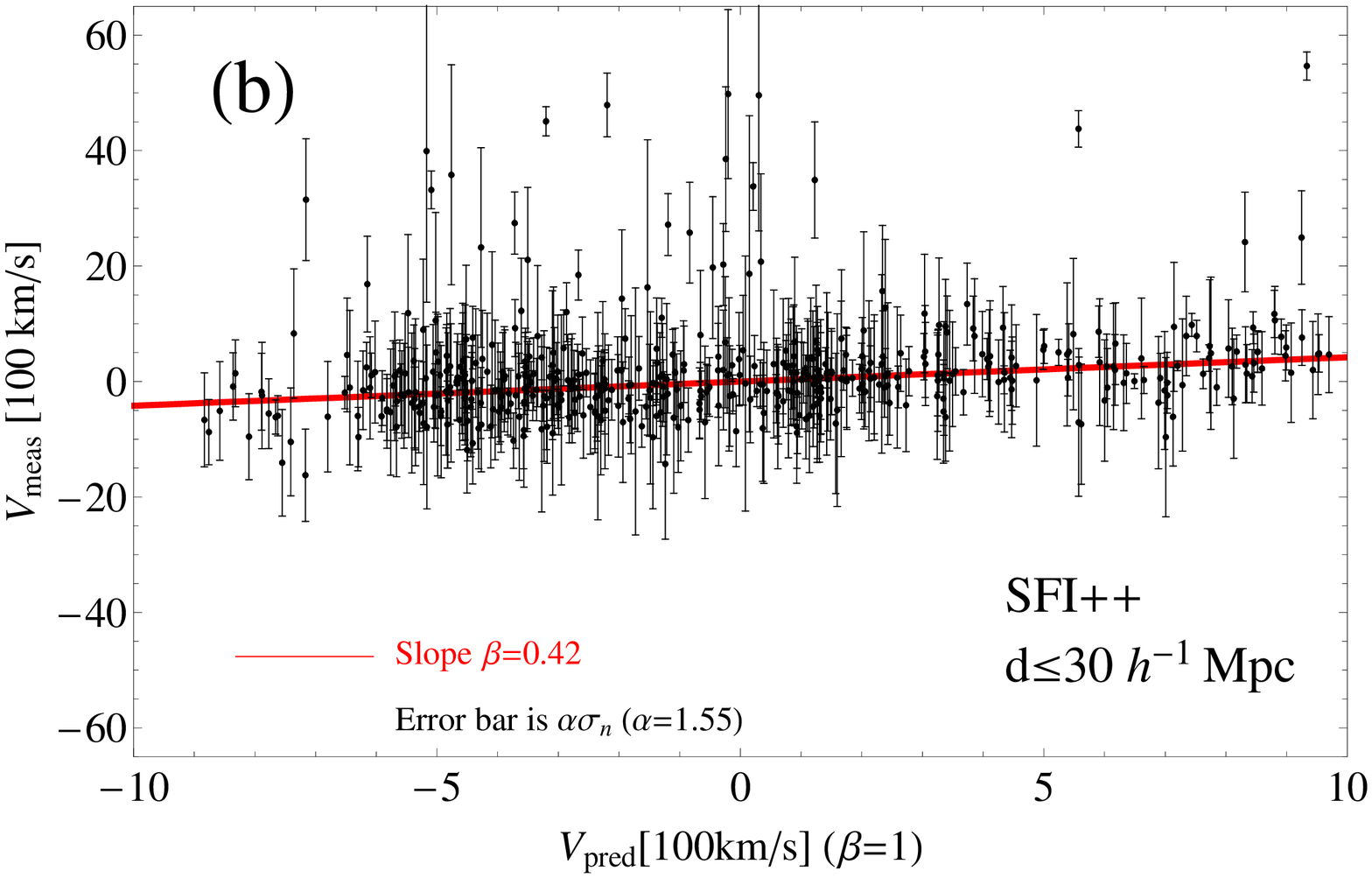}}
\caption{Panel (a): Posterior distribution of the linear
flow parameter $\beta$ ($\equiv f/b$) for several surveys, together with a
combined ``hyper-parameter'' analysis.  Panel (b): comparison between measured
peculiar velocities (from the SFI++ catalogue) and the smoothed reconstructed
velocities (from the PSC$z$ galaxy redshift survey).
Figure taken from \citet{Ma12a}.} \label{fig:vvfig}
\end{figure*}

Equation~(\ref{eq:cz}) suggests that, in order to calculate the
peculiar motion of a galaxy, one needs independently to measure
the cosmological redshift, $z$, and the real distance, $r$ (in the
same comoving frame). To determine the redshift, one uses
spectroscopy, comparing features in the galaxy spectrum with known
features in terrestrial spectra.  Spectroscopic techniques are
sufficiently mature that they produce negligible measurement
error.  On the other hand, determining the distance of a galaxy is
considerably more difficult, because one needs to use an empirical
relation between two intrinsic properties of a galaxy (or object
within a galaxy, such as a supernova) to infer the distance. Among
the most well used methods are the Tully-Fisher (hereafter TF)
relation \citep{Tully77} for spiral galaxies, the Fundamental
Plane for elliptical galaxies \citep{Faber76}, and the
luminosity-distance relation for Type-Ia
supernovae~\citep{Tonry03}. Such distance estimates normally have
much larger errors than spectroscopic redshifts (for more
information on distance indicators, please see
Appendix~\ref{app:dist}).

In Fig.~\ref{fig:sfifig}, we show the largest and densest peculiar
velocity catalogue currently available.  This is a close to full-sky sample
of $3456$ spiral galaxies in the field and in galaxy groups.
One can see that some peculiar velocities are positive, i.e.\ the galaxies
are moving away from us (coloured red), while others are moving towards us
(blue).  The question is, what determines the motion of
these galaxies?

In the standard $\Lambda$ cold dark matter (hereafter
$\Lambda$CDM) cosmology, gravitational instability causes the
growth of density perturbations and the emergence of the cosmic flow
field.  In the regime where the density perturbation is
linear, the time evolution of the density contrast
$\delta\equiv(\rho-\overline{\rho})/\rho$ and the galaxy peculiar
velocity field are linked through the continuity equation
\citep{Peebles93}:
\begin{equation}
 \vec{v}_{\rm g}(\vec{x})=\frac{H_{0}f_{0}}{4 \pi }
 \int {\rm d}^{3}\vec{x}^{\prime }
 \delta _{\textrm{m}}(\vec{x}^{\prime },t_0)
 \frac{(\vec{x}^{\prime }-\vec{x})}{\left\vert \vec{x}^{\prime }
  -\vec{x}\right\vert ^{3}},
\label{eq:vg0}
\end{equation}
where $f_{0}\equiv{\rm d}\log(D(a))/{\rm d}\log(a)|_{a=1}$ is the
present day growth rate, with $D(a)$ the growth factor describing
how fast the density contrast grows.  A complication is that
galaxies and matter are related by some kind of bias parameter
$b$, such that $\delta_{\rm g}=b \delta_{\rm m}$. This means that
the relationship between $\vec{v}_{\rm g}$ and $\delta _{\rm m}$
is governed by the combination $f_0/b$, normally labelled $\beta$.
Equation~(\ref{eq:vg0}) can be adapted to provide at least two
different kinds of test.
\begin{enumerate}
\item The velocity field ($v_{\rm g}$) is completely predictable
given the underlying density distribution ($\delta_{\rm m}$). If
we can measure both $v_{\rm g}$ and $\delta_{\rm m}$, we therefore can check
the predictions of linear perturbation theory, and constrain cosmological
parameters.
\item We can use Eq.~(\ref{eq:vg0}) to compute the overall
``streaming motion'' of galaxies within a given volume.  This can be
compared with $\Lambda$CDM predictions of cosmic bulk motion as a function of
scale and also used to test homogeneity.
\end{enumerate}

The twin approaches, of comparing $\delta$ vs.\ $\vec{v}_{\rm g}$
and determining the bulk motion, have dominated this research area
for over two decades (e.g.\ see previous reviews
\citealt{Dekel94a,Zaroubi02}). As a recent example, \cite{Ma12a}
used the PSC$z$ (Point Source Catalogue of redshifts) sample
\citep{Fisher1995,Branchini99,Saunders00} to trace the underlying
mass density field within $300\hmpc$, under the assumption of
linear and deterministic bias.  Then using this $\delta_{\rm
m}(\vec{x})$ function the predicted velocity field can be
compared with the measured velocities.  By minimizing the
deviations between the smoothed reconstructed velocities and the
measured velocity catalogues (including the Early galaxy catalogue
ENEAR, \citealt{Bernardi02}, Spiral galaxies in SFI++
\citealt{Springob07} and two supernova catalogues, \citealt{Tonry03}
and \citealt{Turnbull12}), constraints were placed on $\beta$,
as shown in Fig.~\ref{fig:vvfig}a. Different peculiar velocity
catalogues give quite consistent constraints on $\beta$, and the
velocity--velocity comparison is consistent with the predictions
of $\Lambda$CDM (Fig.~\ref{fig:vvfig}b).  More detailed future
comparisons between these two vector fields would then constitute
tests of the gravitational instability paradigm and for the
$\Lambda$CDM model (e.g.\ see the constraint on modified gravity models
in \citealt{Hudson12}).

\section{How large is the streaming motion?}
\label{sec:bulk}

\begin{figure}
\centerline{\includegraphics[bb=12 22 497
378,width=3.0in]{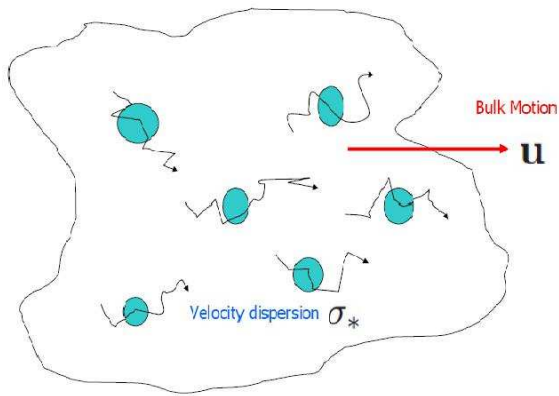}}
 \caption{A schematic illustration of bulk motion $\vec{u}$ and small
scale velocity dispersion $\sigma_\ast$.}
 \label{fig:bulk}
\end{figure}

\begin{figure*}
\centerline{\includegraphics[bb=0 0 492
340,width=3.4in]{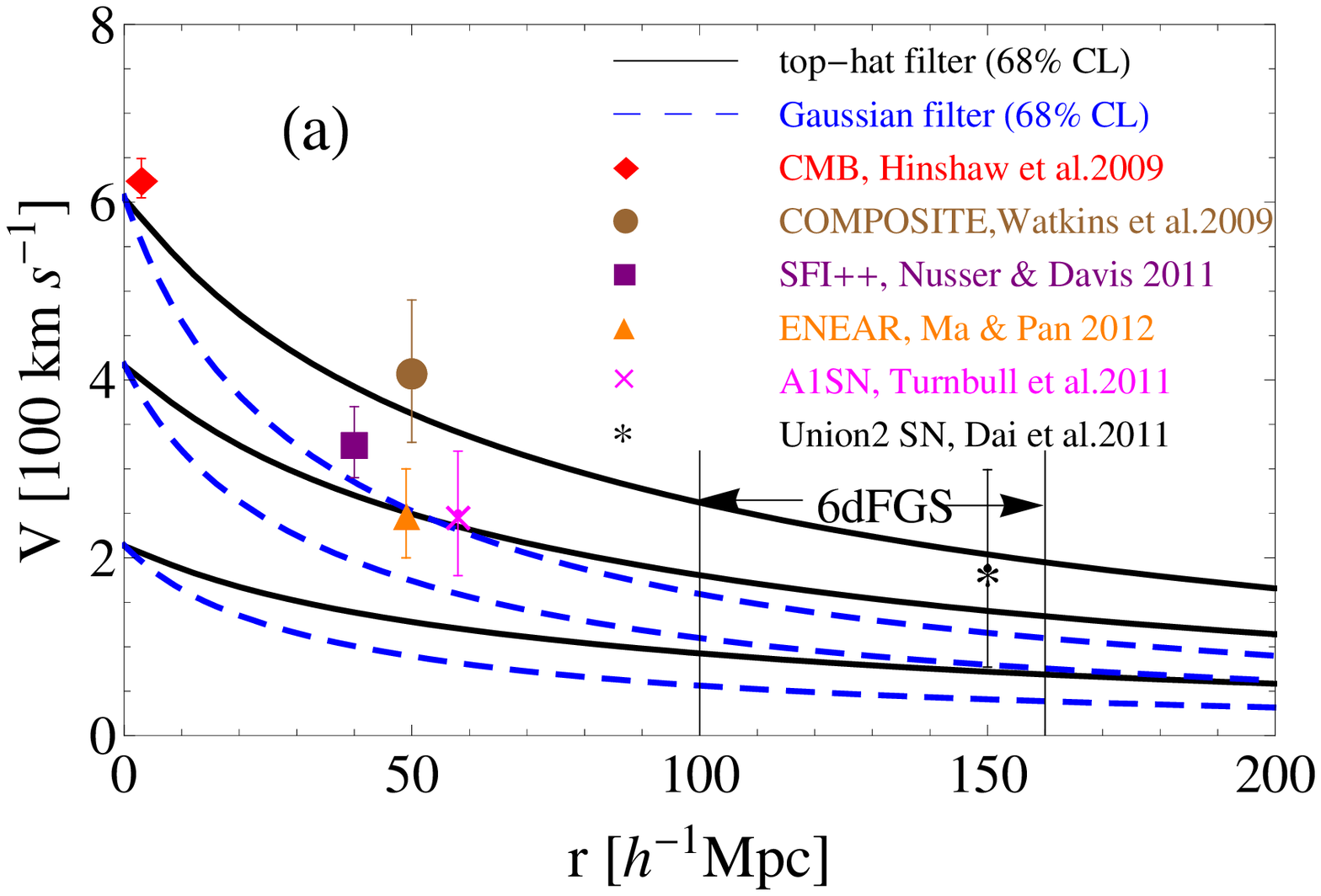}
\includegraphics[bb=0 -70 563 296,width=3.4in]{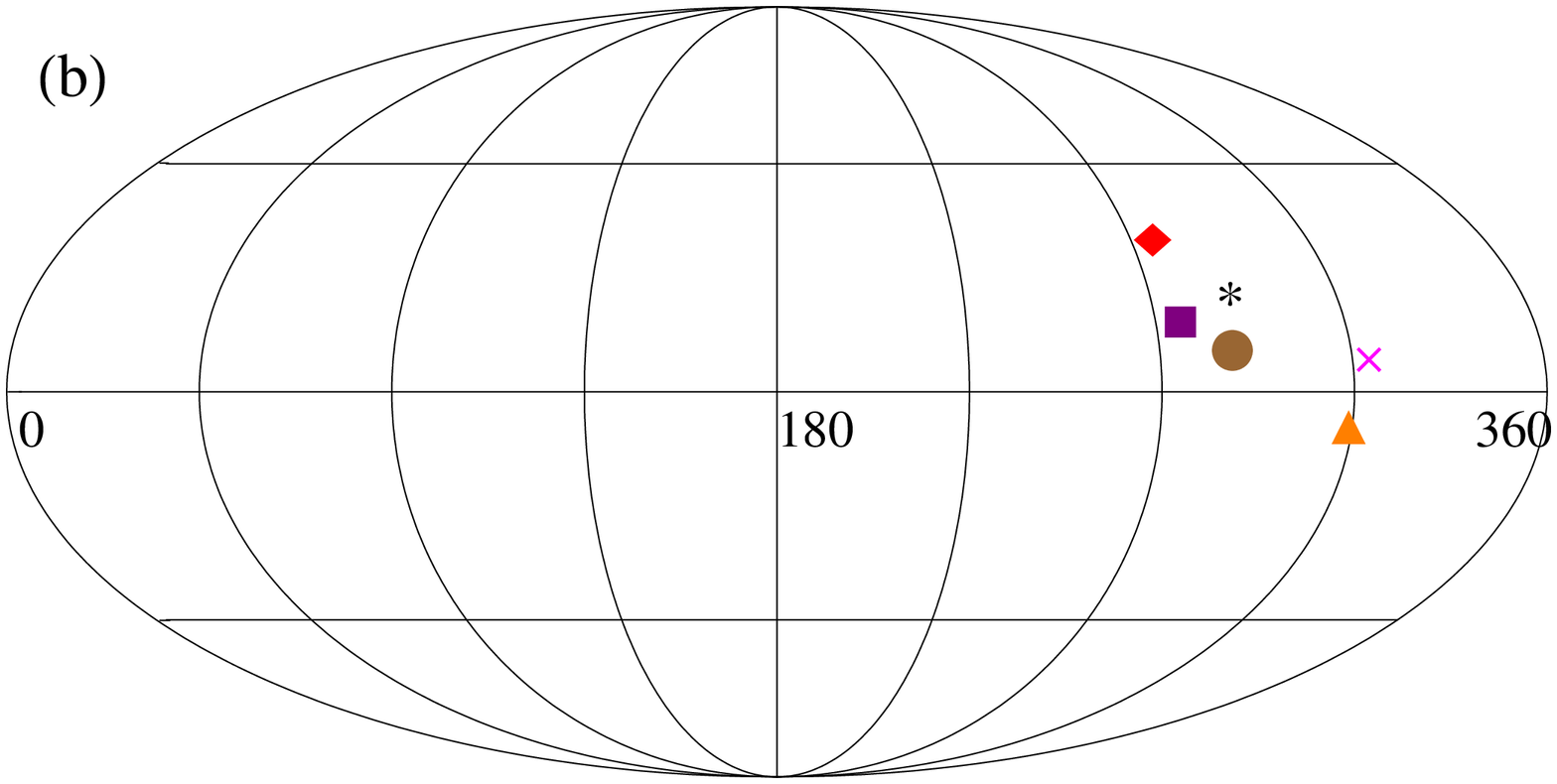}}
 \caption{Panel (a): comparison between the bulk velocity amplitude and
 theoretical evaluation (Eq.~(\ref{pveq})) for various catalogues and
 methods at difference depths.  The theoretical expectation is given for two
 different choices of window function (top-hat in solid black and Gaussian in
 dashed blue), together with the $\pm1\,\sigma$ expected scatter.
 Panel (b): direction of the bulk flow found using different data sets.}
 \label{fig:Constraints}
\end{figure*}

The above comparison tells us the extent to which
the velocity field follows the mass-density field in the way predicted by
the standard cosmology.  The individual velocity
modes, on the other hand, contain information both on large and
small scales. To see this, consider a few galaxy samples on
the sky (as shown schematically in Fig.~\ref{fig:bulk}), which may
share the same streaming motion towards some particular direction,
even while each of them has its own small-scale velocity.
The bulk motion reflects the large-scale density perturbations,
while the small-scale peculiar motions can be considered as a velocity
dispersion arising from the local environment.

The straightforward way of calculating the bulk flow is to average all
of the peculiar velocities out to survey depth $R$ as
\citep{Nusser11a}
\begin{eqnarray}
\vec{V}_{\rm{bulk}}(R)=\int_{r<R} \vec{v}_{\rm{pec}}(\vec{x}) {\rm d}^{3}x
 \left/\int_{r<R} {\rm d}^{3}x. \right.\label{eq:bulkdef}
\end{eqnarray}
However, since individual peculiar velocities
$\vec{v}_{\rm{pec}}(\vec{x})$ are subject to both an intrinsic
velocity dispersion ($\sigma_{\ast}$) and measurement error
($\sigma_{\rm n}$), it is hard to determine the left-hand-side of
Eq.~(\ref{eq:bulkdef}). Instead, a variety of methods have
been developed to estimate the bulk flow given the measured radial
velocities. Figure~\ref{fig:Constraints} shows the results of some
bulk flow reconstructions on various scales, including: the
observed CMB dipole that characterises our peculiar motion with
respect to the CMB rest frame \citep{Hinshaw09}; a ``minimal
weight'' method for a composite galaxy catalogue as well as the Type-Ia
supernovae catalogue (\citealt{Watkins09} and \citealt{Turnbull12}); the
inverse Tully-Fisher relation for the SFI++ catalogue \citep{Nusser11a};
and the Fundamental Plane method for the ENEAR catalogue
\citep{Ma12c}.

What is the amplitude of the bulk flow that we expect in the $\Lambda$CDM
model and how certainly can it be predicted?  The expected variance $\sigma_V$
of the bulk flow on a given scale can be calculated knowing the power spectrum
of density perturbations and using Eqs.~\ref{eq:vg0} and \ref{eq:bulkdef}.
The probability distribution
of the bulk flow is a Maxwell-Boltzmann distribution, skewed towards higher
velocities \citep{Bahcall94,Coles02}:
\begin{eqnarray}
p(V){\rm d}V 
=\sqrt{\frac{54}{\pi }}\left( \frac{V}{\sigma _{V}}\right)^{2}
 e^{-3/2(V/\sigma_{V})^2}
 \left( \frac{{\rm d} V}{\sigma _{V}}\right).
\label{pveq}
\end{eqnarray}%
The variance of the flow magnitude is also
asymmetric, because of the non-Gaussian shape for the distribution.

Fig.~\ref{fig:Constraints}a shows the most likely value and
$\pm1\,\sigma$ scatter of the bulk flow as a function of scale
$R$, using two choices of smoothing function (top-hat and
Gaussian).  One can see that the size of bulk motion decreases
with $R$, which is because more modes are averaged over in a
larger volume.  In addition, the scatter is larger for smaller
scales, because of the increased sampling variance.  It is also
clear that the expected scatter means that bulk flow measurements
can never be very strongly constraining.  Nevertheless, they
provide good tests of the standard picture. We certainly expect
that for very large scales the Universe becomes homogeneous, and
so as $R \rightarrow \infty $ the bulk velocity should be close to
zero in the $\Lambda$CDM model.  However, deviations from that
model could give quite different predictions, e.g.\ if there were
isocurvature-type initial perturbations, then it would be possible
to have a large relative motion between the matter and CMB rest
frames \citep{Turner91,Ma11a}.

The observational estimates plotted on Fig.~\ref{fig:Constraints}
converge with the theoretical prediction out to $150\hmpc$, but at
larger distance the situation is still not entirely clear.
\citet{Kashlinsky11} has argued that there is a ``dark flow'' with
an amplitude around $1000\,{\rm km}{\rm s^{-1}}$ on scales of $z
\simeq 0.2$, while recent results from the \textit{Planck\/}
satellite did not find such a flow \citep{Planck13}.  Further
information on large-scale bulk flows will come from the
six-Degree-Field Galaxy Survey (6dFGS), which is probing the
southern hemisphere out to $150\hmpc$ with the largest peculiar
velocity survey yet constructed.

\section{Tracing the missing baryons}
\label{sec:missing-baryons}

\begin{figure*}
\centerline{\vbox{\hbox{
 \includegraphics[bb=14 14 624 472,width=3.5in]{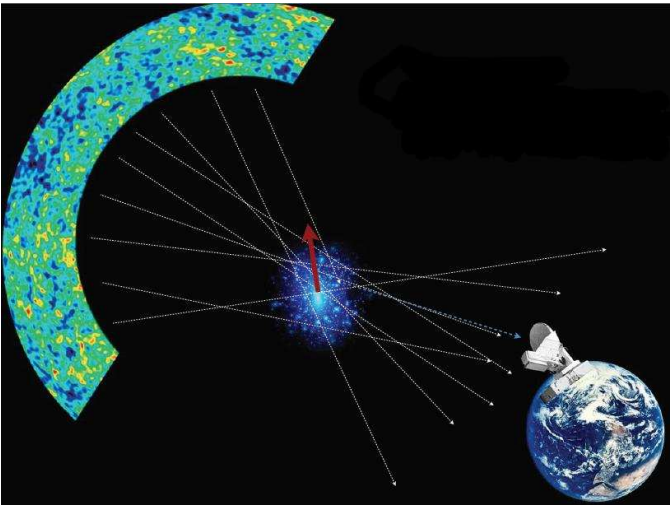}
 \vspace{0.2in}}}
 \hspace{0.1in}
 \includegraphics[bb=14 30 257 257,width=3.2in]{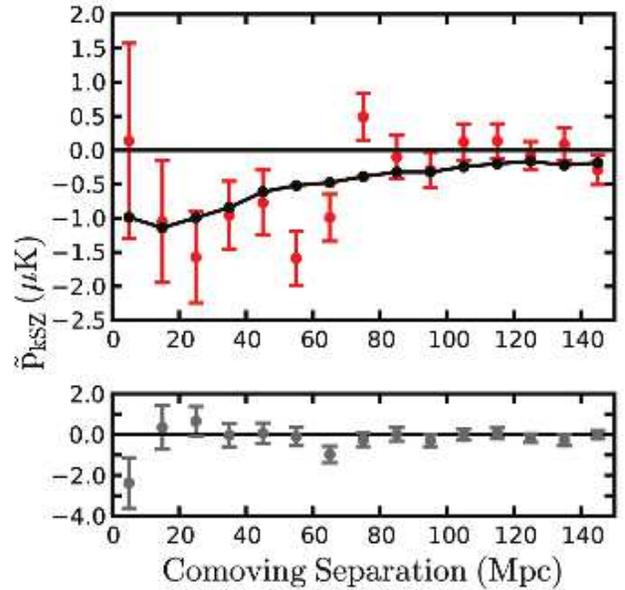}}
 \hspace{0.1in}
 \caption{Left: illustration of the kinetic
 Sunyaev-Zeldovich (kSZ) effect.  A small fraction of the CMB photons are
 scatter by hot gas in clusters of galaxies (and other structures), and
 Doppler shifted by the line of sight velocity of the scattering gas.
 Right: first statistical detection of the kSZ effect from the Atacama
 Cosmology Telescope (ACT), produced by averaging over the positions of
 galaxies.  The red points are data, the black points are the expectation
 from numerical simulations \citep{Sehgal10}
 and the lower panel is a null test \citep[taken from][]{Hand12}.}
 \label{fig:kSZ}
\end{figure*}

For traditional ways of estimating distance, the
error increases as the objects get farther away, and hence it is
increasingly difficult to estimate peculiar velocities for deeper
samples and larger volumes.  However, there is another entirely different
method of estimating the line-of-sight peculiar velocity, which does not
suffer from this distance-dependent uncertainty.  This method is called the
``kinetic Sunyaev-Zeldovich effect,'' as first proposed by
Sunyaev and Zeldovich in 1972 \citep{Sunyaev72,Sunyaev80}. The
idea is shown in the left panel of Fig.~\ref{fig:kSZ}.  CMB photons
travel to us from the last-scattering surface at $z \simeq 1100$,
passing through all the intervening structures.
Ionised gas in galaxy clusters (for example) can re-scatter a fraction of the
CMB photons, Doppler shifting them to either higher or lower energy, depending
on the line-of-sight peculiar velocity of the cluster.  This causes an
anisotropy on the CMB sky of amplitude
\begin{equation}
\frac{\delta T}{T}= \sigma_{\rm T} \int n_{\rm e}
\left(\frac{v}{c} \right) {\rm d}l, \label{eq:ksz}
\end{equation}
where $n_{\rm e}$ is the electron density in the cluster, $v$ is
the radial peculiar motion, $\sigma_{\rm T}$ is the cross-section
of Compton scattering and $l$ integrates through the cluster. The
effect is weak for an individual cluster, and has the same
spectrum as the CMB itself (unlike the larger ``thermal SZ
effect''), making it difficult to extract.  However, by knowing
the positions of large-scale structures it is possible to measure
an average effect. Upper limits from {\it Planck\/} have already
placed constraints on the large-scale monopole and dipole, ruling
out some speculative models \citep{Planck13}. Improvements in
measurement techniques should allow interesting constraints to be
placed on the peculiar motion $v$ of individual galaxy clusters.

In fact the kSZ effect was firstly detected statistically by \citet{Hand12},
who explored the CMB temperature map from the Atacama Cosmology
Telescope (ACT) at the positions of $5000$ Sloan Digital Sky Survey
samples \citep{Gunn06}.  By estimating the velocity differences
between pairs of galaxies with separation distance $r$, and weighting the
ACT map appropriately, \cite{Hand12} were able to detect the average kSZ
effect as a function of $r$ (shown in the right panel of Fig.~\ref{fig:kSZ}).

The kSZ signal is not dominated by the very cores of clusters and hence
is a direct probe of the ``missing baryons'' in the Universe.
As assessed at low redshift by \citet{Fukugita04}, for example,
the baryon fraction from identified sources --
main sequence stars, stellar remnants,
substellar objects and gas -- adds up to only around 10\%
of the amount baryons known from cosmological studies.
Theoretical considerations suggest that most of the undetected
baryons are in gaseous form in the outskirts of virialised groups and
clusters, and in the overdense filaments that connect them.
Numerical simulations \citep{Cen06,Bregman07} show that most of the baryons
are effectively hiding in a phase with temperatures of
$10^{5}$--$10^{7}\,$K (that is difficult to detect directly), in the
Warm-Hot-Intergalactic-Medium (WHIM), which has overdensities in the range
1--1000.  The kSZ effect essentially uses the CMB as a backlight
to find those missing baryons, probing the
outskirts of galaxy clusters and large-scale structure in general.
Further study of cross-correlations between tracers of the gravitational
potential and the gas \citep[e.g.][]{vWHM} hold great promise for teaching us
more about the distribution of gas in the WHIM
on larger scales than has previously been possible to probe.

\section{Analysis tools}
\label{sec:analysis}

In order to use velocities to test cosmological paradigms and constrain
parameters, it is necessary to develop analysis approaches that
carefully control systematic uncertainties.  Many different techniques have
been introduced over the last 30 years.  However, there are some signs that
more recent methods might be offering distinct improvements.  Cosmology is
now dominated by Bayesian approaches to data analysis, with techniques derived
from linear algebra being commonplace, so that likelihood functions and
de-correlation studies become feasible.  Such methods can be applied to
compare velocity and density fields and examine bulk flows.

\cite{Burkey} discuss such approaches for reconstructing the peculiar
veolocity power spectrum.  \cite{Watkins09} describe how to combine multiple
data sets using a minimum variance weighting scheme for constructing the
bulk flow.  \cite{Davis11} use basis functions and the TF relation to compare
velocity and density fields, as well as to calibrate uncertainties
and hence determine $\beta$.  \cite{Ma12a}
discuss the use of a ``hyper-parameter'' technique for combining multiple
noisy peculiar velocity data sets.  These are just a few examples of the
encouraging signs that statistical techniques are keeping pace with the
improving quality of velocity data.

\section{Conclusions}
\label{sec:conclude}

The study of peculiar velocities provides direct tests of several
fundamental issues relating to structure formation.  It is possible to examine
the gravitational instability paradigm, as well as probing homogeneity and
limiting non-Gaussianity.  This is in addition to the use of velocities
for yielding constraints on cosmological parameters, which was once the sole
focus.

In the past the use of peculiar velocities has been hampered by the dominance
of systematic uncertainties in estimates of the peculiar velocity field.
New statistical analysis techniques and improved data, including the use of
the kinetic Sunyaev-Zeldovich effect, mean that there is increasing
interest in the use of peculiar velocities in cosmology.  If this trend
continues, then the next decade may see the promise of cosmic
flows finally being fulfilled.

\vskip 0.1 truein


\appendix
\section{Distance indicators}
\label{app:dist}
The Tully-Fisher (hereafter TF) relation
\citep{Tully77} is an example of a distance estimator for spiral galaxies.
Spirals are flattened systems supported by rotation, with the
rotational velocity well defined due to the flat rotation
curves \citep{Rubin70}.  One expects bigger galaxies to rotate faster,
and empirically there is an approximate power-law relating the luminosity
and rotation speed.  Hence, by measuring linewidths and apparent magnitudes
one can estimate the distance of a galaxy.
An analogous relation exists for elliptical galaxies, connecting luminosity
and velocity dispersion and known as the Faber-Jackson relation
\citep{Faber76}.  An improved correlation for ellipticals comes from
recognising a ``fundamental plane'' in the 3-dimensional space of
surface brightness, velocity dispersion and effective radius.
Tighter distance estimates can be found for galaxies with
Type-Ia supernovae, since they have characteristic light
curves, with the peak luminosity being effectively a constant, so that the
peak apparent magnitude is a distance indicator.

Typically the supernova distances can have a scatter as small as 8\%,
while the best Tully-Fisher and Fundamental Plane distances have
uncertainties of around 20\%.  Improving such distance estimates is crucial
for obtaining better cosmological velocity constraints.  However, more
critical that the statistical scatter is controlling the systematic
uncertainties; distance errors which correlate with galaxy properties or
positions can lead to undesirable biases in velocity analyses.
Hence reducing the systematic uncertainties requires large samples and
analysis techniques which investigate (and correct for) such correlations.


\end{document}